# Multi-hop Moving Relays for IMT-Advanced and Beyond

Ömer Bulakci

Aalto University School of Electrical Engineering, Espoo, Finland

omer.bulakci@ieee.org

*Abstract*— **Relaying is a promising enhancement to current radio technologies, which has been considered in IMT-Advanced candidate technologies such as 3GPP LTE-Advanced and IEEE 802.16m. Relay enhanced networks are expected to fulfill the demanding coverage and capacity requirements in a cost efficient way. Among various relaying architectures multi-hop moving relays can provide additional capacity for the cases when fixed relays are inaccessible or not able to provide adequate solutions in terms of cost. In this paper, an overview of multi-hop moving relays along with some of the envisioned deployment scenarios is presented. Furthermore, different types of multi-hop moving relays are discussed and the challenges are addressed.**

*Index Terms*— **IMT-Advanced; Beyond IMT-Advanced; multi-hop relay; moving relay; relay deployment**

## I. INTRODUCTION

THE expected high data rate transmission with the future wireless communication networks necessitates upgrades for the current network paradigm. In conventional cellular networks, coverage and capacity at the cell border remain relatively small due to low Signal-to-Interference-plus-Noise-Ratio (SINR) where cell-edge users may suffer from large signal attenuation and interference from the neighboring cell transmissions [1]. In addition, there may be a high competition for the available radio resources due to the large number users within the cell. The brute force solution for improving the system capacity and user performance is to significantly increase the density of evolved Node Bs (eNBs), which is also referred to cell-splitting [2]. However, this implies high deployment costs and it is unlikely that the number of subscribers increases at the same rate, which turns out to be economically infeasible for network operators.

A promising solution to overcome the above mentioned problem is deploying relays near the cell edge, which will help to increase the capacity [3][4] or alternatively, to extend the cell coverage area [5][6]. Relays are relatively small nodes with low power consumption, which connect the core network with wireless backhaul through a donor eNB. This feature offers deployment flexibility and eliminates the high costs of a fixed backhaul link (eNB-to-relay link). It is expected that relays lower operational expenditures (OPEX), reduce backhaul costs and enhance network topology [7].

Following the appealing features of relaying, relays have been considered as key network elements to fulfill the requirements of International Mobile Telecommunications Advanced (IMT-Advanced) specified by International Telecommunication Union-Radiocommunication (ITU-R) [8]. Relaying is an essential part of 3rd Generation Partnership Project (3GPP) Long Term Evolution-Advanced (LTE-Advanced) and IEEE 802.16m which are two candidate technologies for IMT-Advanced. Moreover, relaying was investigated during the Wireless World Initiative New Radio (WINNER) project which was a major European Union-funded initiative. The WINNER project contributed to the development and assessment of new mobile network techniques which are already in 3GPP LTE and IEEE 802.16 (WiMAX) standards or under consideration for 3GPP LTE-Advanced and IEEE 802.16m [9]. In addition, very recently in January 2010 another major European Union-funded project, namely Advanced Radio Interface Technologies for 4G Systems (ARTIST4G) has been initiated to improve the allover user experience of cellular mobile radio communications via innovative concepts such as new relay concepts [10].

On relaying researches, there have been two types of relaying architectures: fixed relay node (FRN) and moving relay node (MRN). While FRNs are deployed by the operators in a more deterministic manner, e.g. in coverage holes, MRNs can be deployed flexibly for the cases where FRNs are not available or not economically justifiable. The mobile characteristic of MRN is a degree of freedom that can be exploited to provide coverage. MRN has been investigated in the WINNER [11]-[13] and ARTIST4G [14] projects and is already in IEEE 802.16 standard [15]. It is worth to note that, till Release 10 FRN has been considered in 3GPP LTE-Advanced at least for coverage extension [16] and MRN might be considered in the next releases.

In this paper, an overview of the mobile relaying concept as well as different deployment scenarios is provided. This is followed by the discussion on the challenges of different relay deployments, where some of the possible solutions are also addressed. Finally, a conclusion is given.

## II. MULTI-HOP MOVING RELAY: FROM THEORY TO PRACTICE

### A. Concept of Multi-hop Moving Relays

The use of radio relaying for capacity enhancement and high data rate coverage extension has been discussed in academia for a long time [18]. The earlier studies on relaying were rather theoretical and focused on the network information theory aspect. In [19], Cover and El Gamal formulated capacity theorems for a simple relay channel. Moreover, multiple-input multiple-output (MIMO) techniques for relay networks are also considered and capacity bounds for relay MIMO channels are studied [20].

Multi-hop relaying is one of the appealing relaying technologies considered for the relay networks, where a source node communicates with a destination node either directly or via a relay, not both at the same time. Unless otherwise stated, multi-hop relays are assumed in the rest of the paper. This relaying functionality can be realized either via a fixed relay or moving relay depending on the application scenario. In Fig. 1, the deployments of fixed and moving relays along with some example application scenarios are shown.

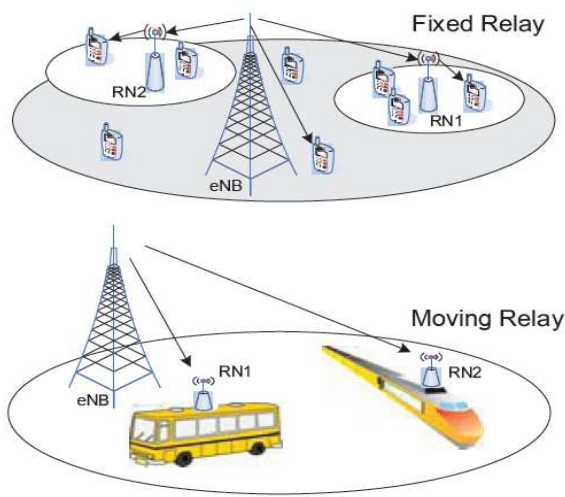

Figure 1. Some application scenarios for fixed and moving relays [21]. A fixed relay can be mounted, e.g. on a lamp post to provide capacity in a hot-spot (RN1) or coverage at the cell edge (RN2). A moving relay can be mounted, e.g. on a bus of train to provide coverage for the users travelling in the vehicles.

*B. The Necessity for Moving Relays*

Fixed relays can be deployed in deterministic areas to provide coverage, e.g. to prevent poor reception because of shadowing. However, there are cases that fixed relays are either unable to provide service at all or are unable to provide economically justifiable solutions [11]. For instance, in case of traffic accidents and football matches the network should cope with an exceptionally high volume of voice or data calls, should guarantee resources for emergency services (e.g. telemedicine) and should be able to provide high capacity during the events. In such unpredicted and/or non-uniform cases a fixed network topology may not provide the necessary service and become overloaded. A possible cost-efficient solution is deploying low-cost moving relays on carriers like ambulances, police cars and buses, which can be switched on a need basis. Additionally, the moving relays can be deployed on the buses to provide coverage on the streets during a certain time of the day (e.g. rush hour), rather than deploying fixed relays. Moreover, moving relays can also provide services for the passengers in a bus, train or riverboat where fixed relays may not be accessible. It is important to note that, the general idea of moving relays is not replacing the fixed relays but rather to become a complementary technology.

*C. Application Scenarios with Practical Challenges and Solutions*

Due to the mobility of moving relays, there can be a large number of usage cases. Furthermore, this feature gives a rise to more severe constraints compared to fixed relays. In this section an overview of the main application scenarios are presented and some of the practical challenges are addressed along with possible solutions.

*1) Type I: Dedicated moving relays serving stationary users*

This type of moving relays is mounted on the transportation vehicles such as buses, trains and riverboats (see Fig. 1) to provide higher throughput and lower handover interruption for the passengers. In this case, the antennas of the moving relay for donor eNB communication are located outside of the vehicle while the antennas of the moving relay for user communication are located inside the vehicle. This feature enables improved channel conditions both for the moving relay and the users since the penetration loss is eliminated. As depicted in Fig. 2, via amplify-and-forward moving relays the call dropping or equivalently radio link failure probability of the users can be significantly decreased [22]. Furthermore, as the users are connected to moving relay instead of eNB, i.e. there is short range link between the users and the access point, the power levels of the transmission are reduced which in turn increases the user battery life time. Additionally, the moving relay is not energy limited like a user.

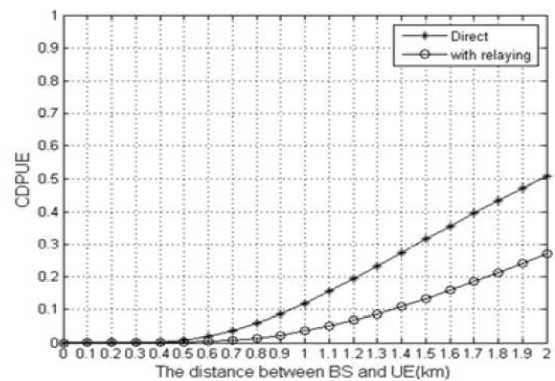

Figure 2. Call Dropping Probability User Equipment (UE) based performance vs. the distance between UE and base station (BS). Via amplify-and-forward moving relay the call dropping probability can be significantly decreased [22].

The challenging issues regarding this type of the moving relays can be stated as follows:

- **Handover (HO)**: In a conventional manner, when a moving relay performs HO, all attached users shall

also perform HO at the same time, which considerably increases the overhead and the latency. This option enables simple moving relay architecture. On the other hand, a moving relay can implement a full set of functions of an eNB such that the HO of the moving relay becomes transparent to its attached users. Although this option decreases the overhead and the latency, it increases the complexity of the moving relay [23].

- **Doppler Shift**: The mobility of the transportation vehicles especially in case of high-speed trains, the received signal may have a Doppler shift associated with it, which is given as:

$$f_D = v \cdot \cos \theta / \lambda,$$

where $\theta$ is the arrival angle of the received signal relative to the direction of motion, $v$ is the receiver velocity towards the transmitter in the direction of the motion and $\lambda$ is the signal wavelength [24]. Therefore, a high Doppler shift can cause loss of the orthogonality and result in inter-carrier interference (ICI). In order to weaken the impact of the Doppler shift, a frequency correction can be provided by the eNB [22]. Note that, this issue is still an open research topic to be addressed.

*2) Type II: Dedicated moving relays serving non-stationary users*

This type of moving relays is also mounted on the transportation vehicles such as buses; however, not to provide service to the passengers rather to provide coverage, e.g. on the streets and in the parks. Such a scenario is shown in Fig. 3, where moving relays are fitted on the public transport buses to provide coverage in Hyde Park in London as denoted by the black circles [12].

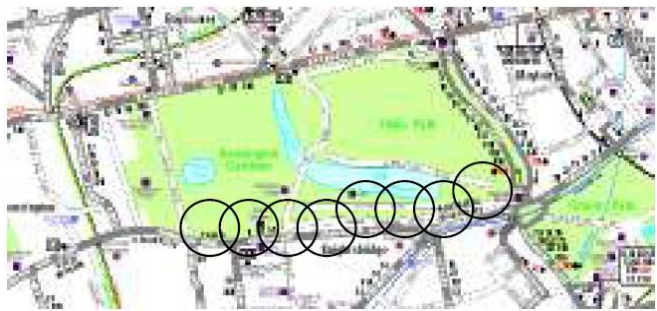

Figure 3. Moving relays fitted on the public transport buses provide coverage in Hyde Park in London [12]. Covered areas are denoted by black circles.

The main issues which should be addressed are as follows:

- **Presence**: The main issue seems to be the not-guaranteed presence of the moving relays [11]. The presence issue mainly defines the quality of the service that a moving relay can provide to out-of-range users. The probability of the presence can be increased using frequent means of public transportation. However, this cannot solve the problem completely.

- **Location**: In order to be able to use specific relay which are in a certain area the location of the moving relays should be known. Moreover, in order to decrease the handover time the path of the moving relay can be exploited to ensure a stable network service. In [25], it is shown that via information based on Global Positioning System (GPS) the latency due the handover can be decreased.

*3) Type III: User terminals as moving relays*

This type of the moving relays are user terminals that can be used by the operators to provide relaying functionalities. Compared to previously mentioned moving relay types, this type promises enhanced system performance without the addition of the costly infrastructure, since the network utilizes the existing user terminals. An example scenario is depicted in Fig. 4.

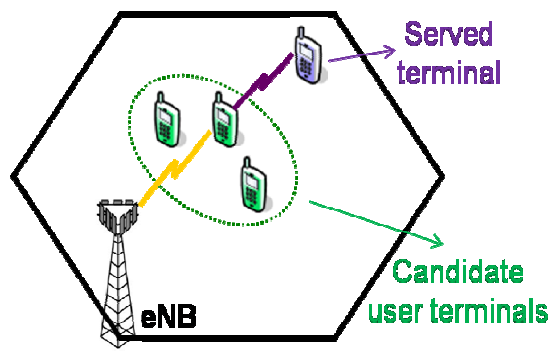

Figure 4. Type III moving relay node is selected out of candidate user terminals. A user terminal is served by one of these candidates.

The utilization of this type poses several challenges:

- **Availability of the candidate user terminals**: The availability of the candidate user terminals is highly unpredictable. A user may not grant this functionality for the reasons of security and fast power drain or this user might just switch off his device. Consequently, another candidate user terminal might be searched. Thus, this scenario is more applicable in hotspot areas with a large concentration of the user terminals [13]. In addition, the operators can provide some incentives to its subscribers if they want to be candidate user terminals.

- **Relay selection overhead**: The signaling overhead and complexity of the relay selection procedure is proportional with the number candidate relays per target user. This number should be reduced while maintaining the performance gains achieved by the relaying. In [26], the set of candidate user terminals is reduced by exploiting the attenuation information and then a low complexity relay selection algorithm is proposed without compromising the performance.

- **Security considerations**: Although the availability of the candidate user terminals can be increased via taken incentives by the operators, the subscribers may not like to use provided services through a user terminal due to security considerations. This issue is particularly important for the business impact of this type of the moving relays. Enhanced encryption techniques should then be investigated to face this challenge.

- **Power limitation of the candidate user terminals**: Including the relaying functionality the battery life time decreases considerably. Hence, the availability of the candidate user terminals might decrease. A possible solution is selecting the candidate terminals according to their capabilities [12]. For instance, a laptop can have higher priority in selection mechanism due to its characteristics such as low mobility, high power availability and high processing capability.

III. CONCLUSION

In this paper, an overview of the multi-hop moving relays is given. The motivation for this enhancement has been discussed for the IMT-Advanced and beyond IMT-Advanced systems. Different application scenarios along with practical challenges and possible solutions are presented.